\documentstyle[epsf,graphicx]{mn}

\def\etal{{et al.\ }}

\def\x2{$\chi^{2}$}

\def\ginga{{\it GINGA }}
\def\asca{{\it ASCA }}
\def\rosat{{\it ROSAT }}

\def\einstein{{\it EINSTEIN }}
\def\chandra{{\it Chandra }}
\def\bepposax{{\it BeppoSAX }}
\def\heao{{\it HEAO }}
\def\gis{{\it GIS }}
\def\sis{{\it SIS }}

\def\x2{$\chi^{2}$}
\def\lunits{$\rm{erg\,s^{-1}}$}
\def\funits{$\rm{erg\,s^{-1}\,cm^{-2}}$}
\def\cunits{$\rm{cm^{-2}}$}
\newbox\grsign \setbox\grsign=\hbox{$>$} \newdimen\grdimen \grdimen=\ht\grsign
\newbox\simlessbox \newbox\simgreatbox \newbox\simpropbox
\setbox\simgreatbox=\hbox{\raise.5ex\hbox{$>$}\llap
     {\lower.5ex\hbox{$\sim$}}}\ht1=\grdimen\dp1=0pt
\setbox\simlessbox=\hbox{\raise.5ex\hbox{$<$}\llap
     {\lower.5ex\hbox{$\sim$}}}\ht2=\grdimen\dp2=0pt
\setbox\simpropbox=\hbox{\raise.5ex\hbox{$\propto$}\llap
     {\lower.5ex\hbox{$\sim$}}}\ht2=\grdimen\dp2=0pt

\begin{document}

\title[ X-ray spectra of Seyfert 2 galaxies] {The X--ray spectra of 
 optically selected Seyfert 2 galaxies. Are there any Sy2
galaxies with no absorption?}

\author[A. Pappa, I. Georgantopoulos, G.C. Stewart and A.L. Zezas ]
       {A. Pappa$^{1}$, I. Georgantopoulos$^{2}$, G.C. Stewart$^{1}$ and A.L.
Zezas$^{1,}\thanks{Current address: Harvard-Smithsonian Center for Astrophysics, 60 Gardes St., Cambridge,
MA 02138, USA }$\\
Department of Physics and Astronomy, University of Leicester, 
Leicester, LE1 7RH \\
Institute of Astronomy \& Astrophysics, 
National Observatory of Athens, Lofos Koufou, Palaia Penteli, 
15236, Athens, Greece \\
}

\maketitle

\label{firstpage}

\begin{abstract}
We present an X-ray spectral analysis of a
sample of 8 bona-fide Seyfert 2 galaxies, selected on the basis of 
their high 
$[OIII]\lambda5007$ flux, from the Ho \etal (1997) spectroscopic 
sample of nearby galaxies. We find that, in general, the X-ray spectra of our
Seyfert 2 galaxies  are complex, with some 
our objects having spectra different from the 'typical' spectrum of 
X-ray selected Seyfert 2 galaxies.
Two (NGC3147 and NGC4698) show no evidence for intrinsic absorption.
We suggest this is due to the fact that when the torus suppresses the intrinsic 
medium and hard energy flux,  underlying emission from the host galaxy,
originating in circumnuclear starbursts, and scattering 
from  warm absorbers  contributes in these energy bands more significantly.
Our \asca data alone cannot discriminate 
whether low absorption objects are Compton-thick AGN 
with a strong scattered component or lack an obscuring torus. 
The most striking example of our low absorption Seyfert 2 is NGC4698.
Its spectrum could be explained by either a dusty warm absorber 
or a lack of broad line clouds so that its
appearance as a Seyfert 2 is intrinsic and not due to absorption. 

%The SED for this particular object shows a deficit in the far-infrared
%emission 
%relative to typical Seyfert 2 galaxies, implying that there is not
%a dusty torus like structure in the central region of the AGN, which
%obscures the broad line region. This intriguing result gives support
%to the \it {unobscured} Seyfert 2 hypothesis.

\end{abstract}


\section{INTRODUCTION}
The discovery of hidden Seyfert 1 nuclei in many Seyfert 2 galaxies
has given much support to the unified theories of Seyferts (e.g
Antonucci \etal for NGC1068). According to the current unification models
both are identical objects which possess a
core-central black hole, accretion disk, broad-line region- and a
thick molecular torus. Objects observed within the opening angle of
the torus are classified as Seyfert 1 objects whereas those seen at
angles intersecting the torus appear as Seyfert 2 sources. Around the
core is ionised gas (warm scatterer) which scatters the primary
emission. 
In some cases nuclear starburst regions have been observed.
It should be pointed
out that the unified theories explain the differences between type 1 and type 2
Seyferts (and in general active galaxies) phenomenologically,
attributing any difference to orientation effects only. 
The X-ray data support the above picture. 
 The X-ray spectra of X-ray selected Seyfert 2 galaxies 
 (eg Turner \& Pounds 1989, Smith \& Done 1996, 
 Turner et al. 1997)  show column densities much
higher than those of Seyfert 1 objects. These columns completely block
the soft X-ray flux but
become transparent to energies $\geq2$ keV. Thus
X-rays prove to be a powerful tool for the study of the
type 2 -obscured- objects, since X-ray photons can penetrate the
obscuring medium and reveal the core to the observer.

Some objects although classified as Seyfert 2 galaxies (e.g. NGC1068
and Circinus, Matt \etal 1997, 1999
respectively) appear to have no excess absorption. It has been shown
that in such cases the direct component in the 2-10 keV range is
suppressed due to the fact that 
the torus is optically thick to Compton scattering, thus the
hard X-ray photons in the 2-10 keV range are 
shifted to lower energies, and finally they
are absorbed after a few scatterings.
In X-ray astronomy terminology these objects are classified as Compton
thick Seyfert 2 galaxies.
In such cases the emission from the nucleus can be inferred from the
photons scattered from the warm scatterer and the inner surface of the
torus itself (cold scatterer).
Therefore Compton thick objects can show low obscuration 
below 10 keV. BeppoSAX observations with the PDS have revealed that a power-law
 emerges through a high column density ($>10^{24} \rm cm^{-2}$) 
above 10 keV in several such cases. 
In such cases the obscuration is large enough to completely block the direct emission leaving a low ``pseudocolumn''
below 10 keV, but small enough to allow transmission above this
energy. However, in some cases the column is so high ($>10^{25} \rm cm^{-2}$)
that the emission is not detected with the PDS.
In addition the observed equivalent width of the $K_\alpha$ iron line increases
as it is measured against a suppressed continuum.
 
Till recently, Seyfert 2 studies were restricted to relatively
X-ray bright Seyfert 2 mainly taken from all-sky X-ray surveys leading to
bias in favour of galaxies with low $N_H$. 
Maiolino \etal (1998) studied a sample of X-ray weak Seyfert 2
galaxies selected by their $[OIII]\lambda5007$ flux and found that the
average obscuration of type 2 AGNs is much higher than that derived
by the former X-ray studies.
This is because the $[OIII]\lambda5007$ flux 
is produced above or below the torus and therefore
can be considered to represent
the central engine and thus provide us with an unbiased sample of
Seyfert 2 galaxies.
Recently Risaliti \etal (1999) studied a large sample of $[OIII]\lambda5007$
selected late-type Seyfert galaxies (Seyfert 1.8, 1.9, 2.0,
$[OIII]\lambda5007>40\times10^{-14}$\funits). The galaxies came from
Maiolino $\&$Rieke (1995), completed with NGC1808. Using X-ray data
from the literature (with the exception of 5 objects, where the
authors analysed the data), they showed that the
average column density for these objects is $N_H=10^{23.5}$, 
with all the Seyfert 2s being obscured by columns with
$N_H>10^{22}$\cunits, giving
strength to the simple model, proposed by the unified theories.
Furthermore they showed that about half are Compton thick
($N_H>10^{24}$\cunits) and they confirmed that intermediate type
1.8-1.9 Seyferts are characterised by an average $N_H$ distribution
lower than that of the genuine Seyfert 2 galaxies.

 Although the above standard model 
 describes very well the spectrum of most Seyferts, 
 recently examples of Seyfert 2 galaxies were found which 
 challenge the unification scenarios. These show no intrinsic
absorption (for example NGC3147, Ptak et al. 1997, NGC7590, 
 Bassani et al. 1999) while  their high 
$f_{HX}/f_{[OIII]}$ ratios are inconsistent 
 with the idea of being Compton thick objects (see Bassani \etal
1999). The peculiar spectra of these Seyfert 2 could be explained,
 for instance,  
 either by the absence of a broad line region or by a high  
dust  to ($\sim$  neutral) gas. In the former case, 
 their appearance as Seyfert 2 is
 intrinsic and not due to absorption.  
The lack of absorbing columns in these galaxies 
 raises important questions about the 
 validity and universality of the standard AGN unification schemes. 
The distribution of the absorbing columns in AGN is also vital
 for models for the synthesis of the XRB (eg Comastri et al. 1995).
It is evident that the range of column densities of the absorbing
material, its structure and geometry remain yet unconstrained 
 and need to be determined with larger samples of galaxies selected 
 in different wavebands. 

In this paper, we present a comprehensive and uniform 
 X-ray analysis with \asca data of  
 8 optically selected Seyfert 2 galaxies from the 
 spectroscopic sample  of Ho et al. (1997). 
 While a few of the objects 
 analysed here have been retrieved from the public \asca
 database,  some are presented here for the first time 
 (NGC1167, NGC2273, NGC3486, NGC4698). 
 Our goal is to explore the validity of the standard model 
 and study the distribution of absorbing columns 
 in a sample with very accurately defined optical properties,
 bona-fide Seyfert 2.
% expanding the previous work by 
% Maiolino \etal (1998) and Bassani \etal (1999)?
This paper is divided as follows: in $\S$2 we introduce our sample; in
$\S$3 we describe our analysis method; in $\S$4 we present the results
of the spectral analysis; in $\S$5 we discuss our results to the
individual objects: in 
$\S$6 we discuss our results; in  $\S$7 we summarise our main
observational results.
%: and in the Appendix, we show the spectrum of
%the serendipitous source in the NGC2273 field.

\section{THE DATA}
We present an analysis of 8 optically selected Seyfert 2  galaxies 
 observed with the
\asca satellite. The list of the data is presented in table 1. 
 Data come from both our own proprietary 
 observations and from the \asca archive.
 Our galaxies are taken from the Ho \etal (1997) spectroscopic sample 
 of nearby galaxies.  This sample contains objects
selected from the Revised Shapley-Ames Catalogue of Bright Galaxies
(RSA; Sandage \& Tammann 1981) with magnitude limit $B_{\tau}=12.5$ mag
in the northern ($\delta>0^o$) sky. As high signal-to-noise and 
 moderate to high resolution optical  
 spectroscopy has been obtained for this sample, very 
 accurate spectroscopic classifications exist for all galaxies. 
 As a consequence we can be confident that all our objects are bona-fide 
 Seyfert 2 galaxies.    
Our eight objects are selected on the basis of  their 
high $[OIII]\lambda5007$ flux. 
Some of our objects (see table 1) have been previously analysed by 
other authors in the hard X-ray band. However, here 
we re-analyse the data in order to present a uniform, comprehensive 
analysis of the brightest $[OIII]\lambda5007$ selected Seyfert 2, for which
X-ray data were/became available, in the Ho \etal sample. 
An $[OIII]\lambda5007$ selected sample should be relatively free from
the selection 
effects and
biases that might appear through X-ray or other optical selection, such as
intrinsic absorption or differences in viewing angle (Ueno \etal 1998).
For 5 of the sources we were able to perform spectral analysis. For
the other 3  (NGC1167, NGC1667 and NGC3486), 
insufficient X-rays were detected for a full spectral analysis and we
restrict ourselves to a hardness ratio analysis. 

\begin{table*}
\caption{The \asca Seyfert 2 sample. The columns contain the following
information: (1) The source name; (2) The sequence number of the
observation; (3)$\&$(4) Optical position of the object; (5) The
exposure time for SIS-0 in ks; (6) The source redshift; (7) The line of sight Galactic hydrogen
column density; (8) References to the previously published hard X-ray data. } 
\begin{tabular}{lccccccc} \hline 

Name     & \asca   & $R.A.^{a}$       & $Dec.^{a}$ & SIS-0   & $z ^{a}$       & $N_H$(Gal)  & References \\ 
         &Sequence &J2000             &J2000       &  exp &              &($\times 10^{21}$\cunits)&   \\ 
(1)     & (2)      & (3)   & (4)    &    (5)             & (6)    & (7)& (8)\\\hline
NGC1167 &77072000& 03 01 42.4       &+35 12 21& 38    &0.016495            & 1.14   & -\\
NGC1667 &71032000& 04 48 37.1       &-06 19 12& 14.5  &0.015167             & 0.55   & 1,2 \\
NGC2273 &74039000& 06 50 08.7       &+60 50 45& 34    &0.006241             & 0.68    & 3\\
NGC3079 &60000000& 10 01 57.8       &+55 40 47& 29    &0.003753             & 0.08    & 4 \\
NGC3147 &60040000& 10 16 53.6       &+73 24 03& 23    &0.009407             & 0.36   & 5   \\
NGC3486 &77074000& 11 00 23.9       &+28 58 30& 41.3  &0.002272             & 0.19   & -\\
NGC4698 &77073000& 12 48 23.0       &+08 29 14& 40.5  &0.003342             & 0.19      & -\\
NGC5194(M51)&60017000 & 13 29 52.35 &+47 11 53.8&30.5 &0.001544          &  0.16      & 6  \\ \hline\
\end{tabular}

NOTE: Units of right ascension are hours, minutes and seconds. Units
of declination are degrees, arcminutes and arcseconds. \\
$^{a}$From the NASA Extragalactic Database (NED).\\
REFERENCES.(1)Ueno \etal 1997; (2)Turner \etal 1997; (3)Maiolino \etal
1998; (4)Ptak \etal 1999; (5)Ptak \etal 1996; (6)Terashima \etal 1998.
\end{table*}

\section{ DATA REDUCTION}
We utilised both \asca \gis and \sis data. We used the standard ``Revision 2'' 
processed data from the Goddard Space Flight Center (GSFC) and data 
reduction was performed using FTOOLS. 
For \gis data we used a circular source region centered on the
source. Background counts were estimated from source-free annuli 
centered on the source cell. Due to calibration differences between
the four SIS chips, we limited our analysis to the on-source chip for
each SIS. We used a circular extraction cell of 3 to 4 arcminutes in
radius. In cases, where the source was centered close to the gap 
between the chips, we followed  the process described in the ASCA ABC
guide (Yaqoob 1997). Background was estimated using rectangular
regions at the source chip, excluding the source. 
%In the case of M51
%where extended emission has been observed in the hard band (Terashima
%\etal 1998) we used a 
%circular source region of 1.5 arcmin in radius in \gis (the same as
%the PSF) in order to minimise the starburst contamination.
%Because in the \sis the source is centered in
%the gap between the chips we restricted our analysis for this
%particular galaxy in the \gis data only.
In the case of NGC2273 where an additional serendipitous
 source close to the galaxy is detected,
we used a circular source region of 1.5 arcmin in radius in both \gis
and \sis in order to minimise any contamination by the nearby source.

\section{SPECTRAL ANALYSIS}

The spectral analysis was carried out using XSPEC v10.
We bin the data so that there are at least 20 counts 
in each bin (source plus background). 
Quoted errors to the best-fitting spectral parameters are 90 per cent
confidence regions for one parameter of interest ($\Delta\chi^2$=2.71).
We  performed spectral fitting, allowing the normalisation for
the \sis and \gis detectors to vary. The fluxes and luminosities given
in the tables are referred to the \gis data. 
Throughout this paper values of $H_o=75\,\rm km~s^{-1}~Mpc^{-1}$ 
 and $q_o$=0.5 are assumed. 
From our analysis, we exclude all  data below 0.8 keV 
 due to uncertainties in the calibration matrices (see George
\etal 1998 for a discussion on this).
We apply relatively simple spectral models, so as to describe the
properties of the whole sample in the context of the unified
model. 
We apply all the models to all the
data sets even though the most complex models may not be required in
some cases.
Where  fits gave us absorption values consistent with the Galactic column 
density or lower  we have fixed the column to the Galactic value.
The latter was determined using the nh ftool, which utilises a map
based on 21-cm measurements and has
resolution of about 0.7 degrees (see Dickey and Lockman, 1990).

\subsection{Building the standard model}

We have used the four major components of the  ``standard model'' in steps:
a) we first apply a single power-law with absorption if required, to all of
our objects, b) we then add a Gaussian line representing the iron $K_\alpha$
emission, c) we add a second power-law component representing scattered
emission; d) finally a Raymond-Smith component at low energies 
 representing a star-forming component is added to the data. 

\begin{figure*}
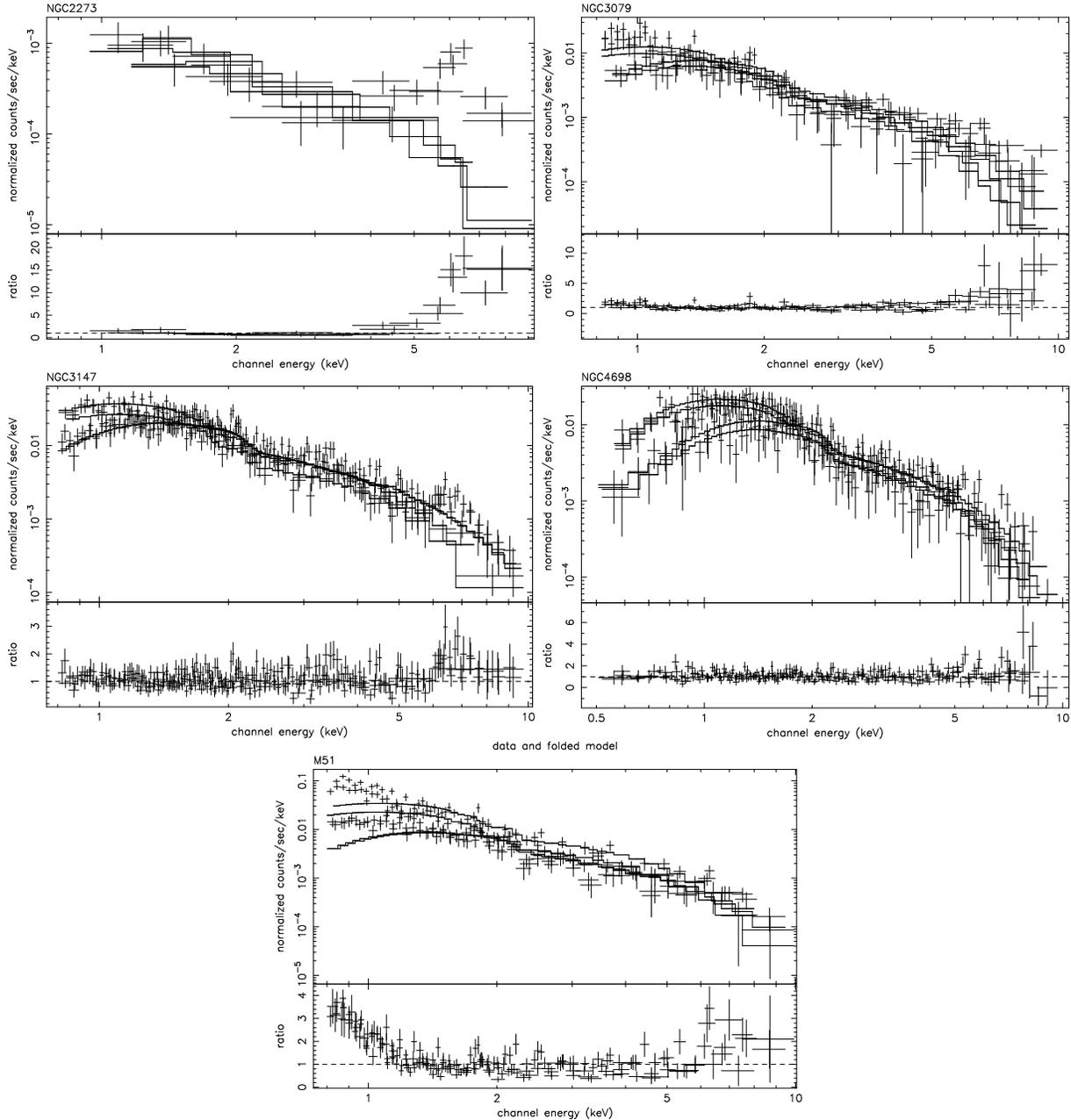

%\rotatebox{270}{\includegraphics[height=8.0cm]{ngc1667.ps}}
\rotatebox{270}{\includegraphics[height=8.0cm]{ngc2273.ps}}
\rotatebox{270}{\includegraphics[height=8.0cm]{ngc3079.ps}}
\rotatebox{270}{\includegraphics[height=8.0cm]{ngc3147.ps}}
\rotatebox{270}{\includegraphics[height=8.0cm]{ngc4698.ps}}
\rotatebox{270}{\includegraphics[height=8.0cm]{po_all.ps}}
\caption{The single power-law model. 
The top panel shows the data with the model and the bottom panel shows the
data/model ratio. In the latter panel the spectral features are clear.}
\end{figure*}

\subsubsection{Single power-law model}
%first utilise the simplest model, a single power-law. 
According to the current theories the primary UV and soft X-ray
photons produced by the disk are Compton scattered by a 
medium  of hot electrons ($\sim$ 50 keV) and reradiated at hard energies.
Such a process can produce 
a power-law with an index of $\Gamma\sim1.9$, in agreement with
observational results, which suggest that the intrinsic X-ray spectrum of 
Seyfert 1 galaxies is well
represented by a power-law with a ``canonical'' spectral index with
$\Gamma\sim1.9$. In the case of Seyfert 2 galaxies, where our line of
sight intersects the torus, we expect the spectrum to show a low
energy cutoff as well.
We thus fit the data with a power-law model $N(E)dE \alpha E^{-\Gamma}dE$,
where N(E) is the photon number density at energy E. The photoelectric
absorption A(E) is parameterised by 
$A(E)=e^{(-N_{H}\times\sigma(E))}$
where $N_{H}$ is the equivalent hydrogen column and $\sigma(E)$ the
photo-electric cross-section taken from  Morrison $\&$ McCammon 1983. 
Figure 1 shows the data and the data/model ratio after this model was
fitted to each object.
Rather unexpectedly this model does provide an acceptable fit in 2
cases, 
where the absorption needed is either the 
Galactic value (NGC3147) or
relatively low (NGC4698). An absorbed power-law model is clearly rejected 
by the NGC2273 and M51 data. The model parameters are shown in table 2. For
the cases of M51, where the obtained spectral slope is unphysically
steep and for NGC2273, where the slope is rather flat, we also fitted the
model with the slope fixed at the nominal value of 
$\Gamma=1.9$, for comparison.

\begin{table*}
\caption{Single power-law model} 
\begin{tabular}{lccccc} \hline 

Name   & $\Gamma$          & $N_H(\times10^{21} \rm cm^{-2})$
&$f_{2-10 keV}^{\, a}$& $L_{2-10 keV}^{\,b}$ &$\chi^2/dof$ \\ \hline
%NGC1667 &$4.37^{+1.99}_{-1.37}$ &0.55g & 0.01 & 0.003 & 34.41/28 \\
%        & 1.9f&0.55g & 0.09 & 0.21 & 45.25/29 \\
NGC2273 & $-0.55^{+0.11}_{-0.11}$ & 0.68g & 0.9 & 0.65& 82.92/22 \\
        & 1.9f & 5.86 & 0.18 & 0.14 & 117.47/22 \\
NGC3079 & $1.93^{+0.12}_{-0.12}$ & 0.08g & 0.74 & 0.20 & 228.06/143 \\
NGC3147 & $1.82^{+0.05}_{-0.06}$ & 0.36g & 1.63 & 2.76 & 289.22/285 \\
NGC4698 & $1.91^{+0.14}_{-0.14}$ & $0.81^{+0.82}_{-0.78}$ & 1.04 &
0.22 & 243.90/250 \\     
M51 & $3.07^{+0.06}_{-0.06}$ &  0.16g &  0.35     &   0.02    & 568.88/178  \\
    & 1.9f &  0.16g &  0.90     &   0.04    & 848.21/179  \\ \hline\
\end{tabular}

NOTE: g indicates that the $N_H$ is set to the Galactic value, 
      whereas f indicates that the parameter value is frozen.\\
     % For all the entries with no associated uncertainties $\chi_{\nu}^2>2$. \\
$^{a}$observed 2-10 keV flux in units of $10^{-12}\rm erg~sec^{-1}cm^{-2}$.\\
$^{b}$unobscured 2-10 keV luminosity in units of $10^{41}\rm
erg~sec^{-1}$, corrected for absorption quoted in column 3. \\

\end{table*}

\subsubsection{Iron $K_\alpha$ line}

Strong evidence supporting the presence of an accretion disk in the
vicinity of a black hole has been given by the detection of an
asymmetric broad emission line at a rest energy of $\sim6.4$ keV in the
Seyfert 1 galaxy MCG-6-30-15.
This is interpreted as $K_\alpha$ iron emission
originating by fluorescence in the very inner parts of an accretion
disk, $3R_S\leq R\leq 10R_S$, about a massive black hole of
Schwarzschild radius $R_S$ (Tanaka \etal 1995). 
The profiles and intensities of the iron lines are expected to be complex
(asymmetric, double peaked) but the sensitivity of the current detectors
cannot provide us with such detailed profiles for the majority of the
observed objects. In the Seyfert 2s,
reprocessing of the nuclear radiation by the obscuring torus 
may also contribute to the line flux.
The mean equivalent width for a sample of  Seyfert 2 galaxies studied with 
\asca  is $363\pm254$ eV (Gilli \etal 1999). 

Some of our objects show line like residuals in the 6-7 keV energy
range, providing evidence for an iron line, which we parameterise
by a Gaussian line.
For the purposes of our study the width of the line is fixed to
$\sigma$=0.01 keV. This is smaller than the instrumental response
and thus is effectively mono-chromatic.
%constrained the line width $\sigma$ to
%be $\leq1.0keV$. In the cases where the line energy lie outside the
%6-7keV range we freezed it at the expected 6.4keV value, taking into account
%the cosmological redshift for each source.
The addition of an iron line provides a
significantly better fit (at $\geq 99\% $ confidence) for three  of
our objects (NGC2273, NGC3147 and M51) although we note here that
the line equivalent widths determined may be strongly affected by the continuum 
shape and may be unphysical. The energy of the iron line is consistent with that
expected for cold iron for all of our objects in which the line is detected.
No line was detected in the spectra of NGC3079 and  NGC4698, for which  we 
give the 90 per cent upper limits.
%The model parameters for each
%object are presented in the table 3, in cases where the line is not
%detected we show the 90 per cent upper limit of the line equivalent width.
%When we allow the width of the line to vary, we obtain unphysically
%large values. This is ({\bf{HELP!!!}})
%On the other hand, the addition of a broad line to explain the flux
%produced in the disk, does not provide a significantly better
%parametarisation of the data for none of our objects.

\begin{table*}
\caption{Power-law plus iron line} 
\begin{tabular}{lccccccc} \hline 

Name   & $\Gamma$ & $N_H(\times10^{21} \rm cm^{-2})$ & 
%E(keV) & 
EW (eV)  & $f_{2-10 keV}^{\, a}$& $L_{2-10 keV}^{\,b}$ &$\chi^2/dof$ \\ \hline
%NGC1667 &$4.66^{+2.60}_{-0.55}$  & 0.55f& $6.76^{+0.24p}_{-0.76p}$ & $<5000$ & 0.5& 0.03 & 33.09/26 \\ 
NGC2273 & $1.09^{+0.43}_{-0.32}$ & 0.68g 
%& $6.38^{+0.12}_{-0.12}$ 
&
$9080^{+7000}_{-4500}$& 0.54& 0.41& 33.32/20 \\
NGC3079 & $2.02^{+0.13}_{-0.11}$ & 0.08g & 
%$6.93^{+0.07p}_{-0.16}$  & 
$<2600$ & 0.84 & 0.22 & 217.71/141 \\
NGC3147 & $1.80^{+0.10}_{-0.05}$ & 0.36g &
%$6.48^{+0.18}_{-0.14}$ &
 $593^{+288}_{-223}$&1.67 & 2.82 & 267.02/283 \\
NGC4698 &$1.91^{+0.12}_{-0.10}$  & $0.95^{+0.36}_{-0.42}$ & 
%6.4f  &
$<425$ &1.04 &
0.21 & 243.89/248 \\      
M51 & $2.90^{+0.05}_{-0.05}$ &  0.16g & 
%$6.39^{+0.05}_{-0.05}$ &  
$5520^{+1515}_{-1516}$ & 0.53 & 0.02    & 532.33/176  \\ \hline\

%$\sigma (keV)$ & EW (eV)  & $f_{2-10keV}^{\, a}$& $L_{2-10keV}^{\,b}$ &$\chi^2/dof$ \\ \hline
%NGC1667 & 1.9f & 0.55f& 6.3f & 1& 248 & 0.05 & 0.21& 45.25/27 \\ 
%NGC2273 & $1.65^{+0.68}_{-0.48}$ & 0.68f & $6.45^{+0.41}_{-0.16}$ &
%$0.66^{+0.34p}_{-0.25}$ & $32400^{+233600}_{-31400}$ & 0.67 & 0.48& 17.02/19 \\
%NGC3079 & $2.17^{+0.15}_{-0.14}$ & 0.008g & $6.79^{+0.59}_{-0.46}$ &$\geq0.85$ 
%& $7000^{+3000}_{-6500}$ & 0.93 & 0.25 & 199.93/140 \\
%NGC3147 & $1.87^{+0.07}_{-0.07}$ & 0.36g & 6.34f & $\leq0.75$ & $828^{+318}_{-782.4}$&1.67 & 2.82 & 269.47/283 \\
%NGC4698 & 2.09f & $0.14^{+0.05}_{-0.04}$ & 6.38f &
%$\geq0.78$ & $1990^{+1100}_{-1560}$ &1.09 &
%0.23 & 237.79/249 \\      
%M51 & $2.99^{+0.35}_{-0.29}$ &  0.16g & $6.61^{+0.49}_{-0.46}$ &
%$\geq0.71$& $24100^{+13200}_{-11100}$ & 0.55 & 0.02    & 107.78/73  \\ \hline\

\end{tabular} 

NOTE: g indicates that the $N_H$ is set to the Galactic value.\\ 
      %whereas f indicates that the parameter value is frozen. \\
      %For all the entries with no associated uncertainties
%$\chi_{\nu}>2$.\\ 
$^{a}$observed 2-10 keV flux in units of $10^{-12}\rm erg~sec^{-1}cm^{-2}$.\\
$^{b}$unobscured 2-10 keV luminosity in units of $10^{41}\rm
erg~sec^{-1}$, corrected for absorption quoted in column 3. \\
\end{table*}

\subsubsection{The scattering model}
According to the unified model we expect that a fraction of the
primary emission should be Thomson scattered into our 
line of sight by a photoionised medium.  
The scattered spectrum has the same shape as the incident
spectrum. We note that in this work the term ``scattering'' is always
referred to scattering off the warm photoionised medium.
Optical polarimetric observations (Tran \etal 1995) have shown 
 that up to $\sim$ 10 per cent of
the primary emission is scattered.
The above scenario can be modeled using two power-laws with the same
photon index but different normalisations and absorptions.
The results are shown in table 4.
Again the energy of the line is consistent with the expected 6.4 keV. 
For M51  we  note that the
obtained slope is unphysically steep and although the
fit has been improved, the model does not describe the data
adequately.
For NGC4698 although we obtain a good fit, the normalisations of the
two power-law components are comparable, and neither component
requires any absorption below 10 keV. In the case of 
NGC3147 the column density is too high to be properly constrained 
in the \asca energy range. Finally, in the case of NGC2273,
the normalisations of the scattered power-law component is almost two
orders of magnitude lower than that of the primary power-law component,
suggesting $\sim$2 per cent scattered flux, in agreement with what it
is typically found for the obscured Seyfert galaxies (Sy 1.9-2.0).

\begin{table*}
\caption{Scattering model} 
\begin{tabular}{lccccccc} \hline 

Name   & $\Gamma$ & $N_H(\times10^{21} \rm cm^{-2})$ 
%& E(keV) 
& EW (eV) & $f_{2-10 keV}^{\, a}$& $L_{2-10 keV}^{\,b}$ &$\chi^2/dof$ \\ \hline
%NGC1667 &$5.38^{+1.61}_{-1.38}$& $289.7^{+408.3}_{-224.7}$ & 6.4f& $<6100$& 0.06 & 0.06 &28.13/24 \\
NGC2273 & $1.78^{+0.61}_{-0.27}$ & $982^{+196}_{-433}$ & 
%$6.26^{+0.30}_{-0.26p}$&
 $555^{+4000}_{-520}$&1.11 & 7.80 & 14.28/18 \\
NGC3079 & $2.20^{+0.30}_{-0.20}$ & $950^{+525}_{-117}$ 
%& 6.40f
& $<3000$ & 1.13 & 0.21 & 192.18/140 \\
%NGC3147 & $1.86^{+0.07}_{-0.05}$ &$2940^{+3400}_{-1633}$  &
%$6.47^{+0.47}_{-0.25}$ & 
%$715^{+1100}_{-600}$&  1.91 & 40.0 & 263.26/281 \\
%NGC4698 & $1.91^{+0.09}_{-0.11}$ & $0.62^{+0.36}_{-0.40}$ &$6.90^{+0.10p}_{-0.90p}$ & $635^{+3700}_{-510}$ &1.07 &
%0.23 & 240.23/248 \\ 
M51 & $4.40^{+0.36}_{-0.21}$ & $64.7^{+8.8}_{-8.0}$  &
%$6.16^{+0.27}_{-0.10}$& 
$4100^{+2300}_{-2110}$ & 0.60 & 0.06   &
298.81/174

  \\ \hline  
\end{tabular} 

%NGC1667 & 1.9f & 0 & 6.3f & 1.0 & $<70000$ & 0 & 0.05 & 0.02 &45.25/25 \\
%NGC2273 & $1.85^{+0.80}_{-0.45}$ & $603.3^{+597.7}_{-476.6}$  & 6.36f
%& $0.34^{+0.60}_{-0.34p}$ & $2520^{+800}_{-1300}$ & 0.96&83.97 & 3.65 & 13.90/18 \\
%NGC3079 & $3.33^{+0.45}_{-0.20}$ & $40.0^{+14.0}_{-8.0}$ & 6.37f &$\geq0.83$ 
%& $1200^{+1200}_{-600}$ & 0.77& 0.86 & 0.31 & 178.99/139 \\
%NGC3147 & $1.88^{+0.05}_{-0.06}$ &$1955.0^{+3045.0}_{-129.3}$  & 6.34f
%& $\geq0$ & $1090^{+3789}_{-1080}$ & 0.91&  1.91 & 33.81 & 261.39/281 \\
%NGC4698 & $2.08^{+0.21}_{-0.18}$ & $1.3^{+3.3}_{-0.9}$ & 6.38f &
%$\geq0.67$ & $1930^{+1220}_{-1415}$ &0.98 &1.08 &
%0.23 & 237.54/247 \\ 
%M51 & $4.73^{+0.82}_{-0.67}$ & $55.6^{+19.0}_{-11.7}$  &
%$6.63^{+0.48}_{-0.43}$ & $\geq0.80$& $40600^{+36400}_{-21800}$ & 0.999 & 0.59 & 0.04   & 84.43/71  \\ \hline 

$^{a}$observed 2-10 keV flux in units of $10^{-12}\rm erg~sec^{-1}cm^{-2}$.\\
$^{b}$unobscured 2-10 keV luminosity in units of $10^{41}\rm
erg~sec^{-1}$, corrected for absorption quoted in column 3.  \\
\end{table*}

\subsubsection{Composite model} 
Infrared (Maiolino \etal 1995) and X-ray studies of Seyfert 2
galaxies (Turner \etal 1997) show that their host
galaxies tend to have energetic starburst regions. Indeed some of our
objects show line like residuals at soft energies, suggesting thermal
emission arising from hot gas. Thus we utilise an
emission spectrum from hot, diffuse gas (Raymond model in XSPEC) to model the 
starburst component, along with a single power-law to account for
the hard photons. 
The temperature is set to be
$\leq1.0\rm\, keV$, in order to constrain the starburst contribution to the
soft band. 
We fit NGC3079 and M51 with this model because these
galaxies show evidence for thermal emission.
In the case of M51 the thermal component is absorbed by the Galactic
column, whereas in NGC3079 excess absorption is required.
This model gives  a good fit for both galaxies (see table 5).

\begin{table*}
\caption{Composite Model}
\begin{tabular}{cccccccc} \hline
Name  &$\Gamma$ & $N_H(\times10^{21} \rm cm^{-2}) $ & KT & 
%$E_{line}$(keV) &  
EW(eV) & $f_{2-10 keV}^{a}$ & $L_{2-10 keV}^{b}$ & $\chi^2/dof$ \\\hline

%NGC1667 & 1.9f & $<6558$ & $0.85_{-0.26}^{+0.25p}$ &-&-& 0.76&0.61& 28.39/26\\
NGC3079 & $0.79_{-0.26}^{+0.31}$ & $17_{-17}^{+93}$ & & & 0.11 &  0.3 & 156.42/139 \\
& & ${66_{-14}^{+12}}^\star$& $0.69_{-0.09}^{+0.11}$   & & & & \\
M51 &$1.70_{-0.19}^{+0.15}$  & $0_{-0}^{+1.6}$ &
$0.71_{-0.02}^{+0.05}$ & 
%$6.36_{-0.21}^{+0.19}$&
$947_{-528}^{+812}$ & 0.9 & 0.07 & 235.18/173 \\
%M51 & 1.9f & $0.12_{-0.12}^{+0.93}$ & $0.71_{-0.26}^{+0.12}$ & $6.43_{-0.18}^{+0.57p}$ & $0.22_{-0.22p}^{+0.78p}$  &
%$2000_{-2000}^{+6530}$ & 5.5 & 0.025 & 79.8/71 \\

\hline
\end{tabular}

NOTE:%t indicates the column density obscuring the thermal component.
$^\star$this value is referred to the column obscuring the thermal
component. \\
$^{a}$observed 2-10 keV flux in units of $10^{-12}\rm erg~sec^{-1}cm^{-2}$.\\
$^{b}$unobscured 2-10 keV luminosity in units of $10^{41}\rm
erg~sec^{-1}$, corrected for absorption quoted in column 3.  \\
\end{table*}

\subsection{The Compton thick models}
If the column density exceeds $\sim10^{24}\rm cm^{-2}$, the obscuring
medium is optically thick to Compton scattering 
and thus the primary emission is suppressed and we only observe the
scattered emission from either the warm matter or the inner surface of
the torus itself. For the latter we use the term reflection. 
Here we will consider the case where the Compton reflection from the inner surface of the torus
dominates the observed emission in the 2-10 keV energy band. 
We therefore
utilise the pexrav model in XSPEC, which describes reflection
occurring from a slab of neutral material subtending a solid angle of
$2\pi$ sr to an X-ray point source located above the slab,
to account for the total 
hard X-ray emission and a power-law to represent the soft
emission. 
For the purpose of this study we consider that the slope of
the underlying power-law spectrum is 1.9 and Galactic absorption.
The normalisation of the reflection is given by R=$\Omega/2\pi$, where
$\Omega$ is the solid angle subtending by the reflector. 
However, as we cannot measure directly the intensity of the intrinsic
power-law which is being reflected, the value of R we obtain is not
physically meaningful.
Therefore we define the ratio A=R$\times f_{(2-10 keV)}/f_{(2-10 keV)sc}$ as an
indicator of the fractional contribution that would have been made to
our observed spectrum after correction for absorption by the pexrav reflector.  
$f_{(2-10 keV)}$ is the flux that would have been produced by the
underlying spectrum in the 2-10 keV band assuming that the emission is
not absorbed by the torus
%this corrected  flux in the 2-10 keV 
whereas $f_{(2-10
keV)sc}$ is the observed flux of the fitted scattered component  in the same band.
Given that typically the  scattering  accounts for $\sim$ 1 per
cent of the total X-ray emission, any value of A that is less than 1 suggests we see only a small
portion of the reflector, and small values mean the reflected
contribution to the observed flux is almost insignificant.
We apply this model to  NGC2273, NGC3147 and NGC4698 which show low absorption.
That model provides a good representation of the data for NGC4698
($\chi^2$=245.04 for 250 degrees of freedom),
NGC2273 ($\chi^2$=24.88 for 18 d.o.f.) and NGC3147 ($\chi^2$=267.11 for
282 d.o.f.) (see also table 6).

\begin{table*}
\caption{Compton thick model}
\begin{tabular}{cccccc}
\hline 
Name      &  $\Gamma$ & 
%$E_{line}$(keV) &  
EW(eV)  & A$^{a}$ & 
$f_{2-10 keV}^{b}$ & $\chi^2/dof$ \\\hline

%NGC1667& 1.9f & -& - & $>10^{-8} $ & 0.01 & 0.003 & 45.21/27 \\
NGC2273  & 1.9f & 
%$6.39_{-0.12}^{+0.13}$ & 
$4460_{-3610}^{+900}$ & $3^{+\infty}_{-1.7}$
& 0.62 & 24.88/19 \\
NGC3147 & 1.9f  
%& $6.48_{-0.14}^{+0.21}$ 
& $548_{+387}^{-508}$ & $0.11_{-0.09}^{+0.08}$ 
& 0.02 & 267.11/282  \\
NGC4698 & 1.9f & - &  $0.15^{+0.02}_{-0.01}$ & 0.01 &
245.04/250  \\
\hline
\end{tabular}

$^{a}$the meaning of parameter A is explained in section 4.2. \\
$^{b}$observed 2-10 keV flux in units of $10^{-12}\rm erg~sec^{-1}cm^{-2}$.\\
%$^{b}$unobscured 2-10 keV luminosity in units of $10^{41}\rm erg~sec^{-1}$ \\
\end{table*}

\section{Results on single objects}
In this section we discuss the results of the spectral fits for each
object in our sample individually and compare our result to previous
X-ray studies. 
NGC1167, NGC1667 and NGC3486 did not give sufficient counts for 
a full spectral analysis to be performed.
We did obtain a
$>3\sigma$ detection in the 2-10 keV band for both objects.
Clues for the spectral shape of the sources come from their hardness
ratio. Here we define hardness ratio as (h-s)/(h+s), where h and s are
the total number counts in the detection cells, in the 2-10 and 1-2
keV bands respectively. For our analysis we chose to use GIS data only.

\subsection{\it NGC1167}
 
The hardness ratio of the
source is -0.22$\pm0.12$, which corresponds to a power-law of $\Gamma=2.5\pm0.3$
assuming Galactic absorption. This corresponds to an observed flux in the
2-10 keV band of
$4.0\times10^{-14}$\funits 
and a luminosity of $2.0\times10^{40}$\lunits. 
We note 
that the derived spectral
index is rather steeper than is typical for Seyfert galaxies, suggesting the
possible presence of an additional soft
component, possibly coming
from a starburst region in the vicinity of the AGN.

The low $f_{HX}/f_{[OIII]}$=0.23 ratio (see section 6 for a detailed
discussion on the implications of this ratio) for this source
would suggest  a Compton thick Seyfert 2 galaxy. On the other
hand the HR analysis showed that NGC1167 has a steep spectrum, while
in the context of a Compton thick interpretation we would expect a flat spectrum.
Unfortunately the HR
analysis provides only an indication for the spectral shape and no
information about multiple components contributing to the spectrum. 
It is quite likely that the X-ray spectrum of NGC1167 is  complex
with different components contributing to different energies.
For example a very strong soft excess, possibly originating from
intense starforming activity, could produce a steep spectrum
even if the emission from the central source is completely blocked by a
Compton thick absorption screen. However,
since NGC1167 is too faint for any further spectral analysis to be
performed, the nature of its X-ray emission cannot be determined.

\subsection{\it NGC1667}
NGC1667 has shown a  decrease of  a
factor of $\sim$150 in the 2-10 keV flux 
between its discovery (e.g. Polleta \etal 1996, Turner
\etal 1997).
The object is too faint to perform any spectral analysis. The hardness
ratio of the source is -0.52$\pm$0.14, which corresponds to a
power-law of $\Gamma=3.2\pm0.4$. Assuming Galactic absorption the
observed 2-10 keV flux is then $\sim8\times10^{-14}$\funits, which
corresponds to a luminosity of $1.3\times10^{41}$\lunits.
The slope is steeper than the canonical for Seyfert galaxies and, again,
some fraction of the emission could be attributed to starburst emission.
Indeed, Radovich \& Rafanelli (1996) find evidence for star formation within
10 kpc of this source, thus favouring the latter interpretation.
%However, the data are not described adequately by a reflection
%dominated Compton thick model and thus  
The low $f_{HX}/f_{[OIII]}$=0.1 ratio for this source
would suggest it as a Compton thick candidate.

The substantial
reduction in the hard X-ray flux over a $\sim 20 $ yr timescale
which has, presumably, not yet been reflected in the narrow-line region
would provide an alternative hypothesis to explain the anomalous ratio.
Indeed, assuming that the 
reduction in the hard X-ray flux over a $\sim 20 $ yr timescale, has
not yet been reflected 
in the narrow-line region, we find that the Narrow Line Region in NGC1667 should
 be located at $>$6 pc. Given that the NLR lies 10-100 pc from the central source
 it appears
that indeed the substantial change in the hard X-ray flux has not
reached the NLR yet. 
This explanation is the most likely explanation of  the low 
$f_{HX}/f_{[OIII]}$ ratio and NGC1667 is  probably Compton thin.

\subsection{\it NGC2273}
This galaxy has been studied by Maiolino \etal  (1998) using 
 {\it BeppoSAX} data.
They found that their data are best fitted by a Compton thick
reflection dominated model. 
In addition they did not detect any emission with the PDS on board 
\bepposax and thus they suggested that if the Compton thick model is valid
the absorbing column in our light of sight must be larger than $10^{25}$
\cunits.
 The ``scattering'' model provides a
better fit to the \asca data (with a column close to $10^{24}$ $\rm cm^{-2}$ and 
 an equivalent width of $\sim$500 eV) and thus  is our preferred fit.
It is apparent from figure 1 that the flat spectrum originally obtained from a simple 
power-law fit is caused by a turn-up in the spectrum at energies above $\sim 5$keV.
Indeed a spectral fit over the restricted 0.2-5 keV range give a good fit to
power-law with a slope of $1.5\pm 0.6$ consistent with the canonical AGN and
our scattering interpretation.   
The ratio of the unobscured
hard X-ray emission  to the [OIII]$\lambda$5007 emission is  low, indicating that we do not observe the
primary emission and  the Compton thick model described in the text, yields a poor fit.
%In addition it should be noted that when we fit the data with a single
%power-law model plus an iron line, we obtain a flat index (albeit with
%a big scatter), and the equivalent width of the line is too large to be
%explained by a Compton thin model. 
It is likely that the spectrum of
NGC2273 is more complex. Probably a warm absorber medium and/or
starburst emission contribute to the soft X-ray spectrum of this
Seyfert 2, whereas the emission line around 6.4 keV could be a blend
of iron lines. Unfortunately the low quality of these data do not
allow us to separate the components, which may contribute to the NGC2273
X-ray spectrum.

\subsection{\it NGC3079}

The \einstein IPC detected NGC3079 at the 3.2$\sigma$ level (Fabbiano \etal
1982), with a flux of $3.7\times10^{-13}$\funits. The \rosat PSPC data
are dominated by a nuclear point source but emission is detected
up to 2'.5 from the nucleus (Reichert \etal 1994). Pietsch \etal
(1998) resolved the X-ray emission with \rosat PSPC and HRI into three
components. a) extended emission in the innermost region, with
$L_x=3\times10^{40}$\lunits, coincidents with the super-bubble seen in
optical images, b) emission from the disk of the galaxy that can
partly resolved in 3 point-like sources and c) very soft X-shaped
emission from the halo extending to a diameter of 27 kpc.
Ptak et al. (1999) first presented the \asca data for this galaxy. 
 Their best-fit spectral parameters have large 
 uncertainties:  the power-law $\Gamma\sim2.20^{+2.0}_{-1.0}$ 
and $N_H\approx 6_{-5}^{+4}\times10^{21} $\cunits. The starforming
component, was described by a Raymond-Smith model with kT=0.14($<$0.54) keV. 
Our analysis is in broad agreement with Ptak et al. (1999),
 although the spectral index is somewhat flatter. 
 Our best-fit model is the ``composite'' one (see table 5)
 while no significant  iron K$_\alpha$ emission is detected.
 No variability was detected by Ptak et al. (1998) 
 consistent with a Compton thick scenario. 
Indeed the  X-ray to [OIII]$\lambda$5007 ratio supports either a Compton
thick interpretation or a substantial starburst contribution.
Again NGC3079 seems to have a complex spectrum and the nature
of the X-ray emission cannot be determined with the current data.

\subsection{\it NGC3147}

Emission in the vicinity of NGC3147 was detected at $3\sigma$ by \heao (Rephaeli \etal 1995) but 
no significant signal was detected by \einstein, implying a decrease in the flux between
 these two observations.
Ptak \etal (1996) first studied this object with \asca. They found
that the data are well fitted by a simple power-law with $\Gamma$=1.9 
and there is no indication for absorption. 
Our results are in excellent agreement with those 
 of Ptak et al. (1996). Our best-fit model is the single power-law with the iron line.
 In principle, other more complicated models 
 such as the scattering model or the Compton thick model 
 provide equally acceptable $\chi^2$. 
  In the case of the ``scattering'' model we obtain 
 a very large column of $10^{24}$ \cunits  
while the scattered emission is   $\sim$ 5 per cent of the
primary component. 
However, such high columns cannot be probed by \asca, thus we do not
show the model parameters for this model in table 4.
In addition the Compton thick model 
provides an equally good representation of the data statistically;
 here it is the scattering component which dominates the fit.
 The absence of variability (Ptak et al. 1998) could in principle favour 
 such models. 
 However, although the above models provide good fits to the data
 they are rather contrived as the \asca bandpass does not allow
us to constrain any models with an obscuring column 
 higher than $\sim 10^{24}$ \cunits. 
 Indeed,  a single power-law model is identical to 
 a scattering model with $N_H>10^{24}$ \cunits \, in 
 the \asca band as these  
 columns absorb most photons below 10 keV. 
 It is therefore interesting that 
 the [OIII]$\lambda$5007 emission does 
 not favour the above two models (composite and scattering).

\subsection{\it NGC3486}

The hardness ratio
is 0.29$\pm$0.12 . This corresponds to a quite flat power law of 
 $\Gamma=1.2\pm0.3$ assuming Galactic absorption, suggestive of high amounts of obscuration. 
 For example  the well-known nearby Compton thick AGN
 (eg Circinus, NGC1068) 
 exhibit flat spectra below 20 keV due to 
 the combination of the  reflection and the 
 scattering components. 
The observed flux assuming Galactic absorption is
$\sim5\times10^{-14}$\funits.

We carried out simulations in order to determine the amount of
obscuring medium needed to obtain a change of $\Gamma$ from 1.9, which
is the common value for AGNs to 1.2. We found that the source should
be obscured by a column density of $N_H=3.2\times10^{21}$\cunits. 
Using $\Gamma=1.9$ and $N_H=3.2\times10^{21}$\cunits we obtain an
observed flux of $\sim5\times10^{-14}$\funits, which corresponds to a
luminosity of $\sim5\times10^{38}$\lunits. This is too low for a
Seyfert galaxy, suggesting high absorption in our line of sight
(but see Roberts $\&$ Warwick 2000). 

The [OIII]$\lambda$5007 ratio is relatively low (2.9). In order for
the ratio to be comparable to the ratios observed in Seyfert galaxies
(see section 6)
the source should be obscured by a column density of $\sim$
1$\times10^{24}$ cm$^{-2}$ assuming a power-law of $\Gamma$=1.9.
This is much higher than the column derived from the HR analysis assuming
$\Gamma$=1.9. This indicates that the spectrum cannot represented by a
single power-law model, and that other components contribute to the
X-ray spectrum as well.

\subsection{\it NGC4698}

NGC 4698 was observed with \einstein and its 0.2-4.0 keV flux is
$2.8\times10^{-13}$ \funits. A detailed discussion on this object is
presented in section 6.

\subsection{\it M51}

M51 is known as the 'Whirlpool galaxy'. The \einstein HRI detected
X-ray emission from M51 with a luminosity $L_x=3.0\times10^{40}$\lunits
in the 0.2-4.0 keV band. The emission is extended and the luminosity of
a point source at the nucleus is
$L_{0.2-4.0 keV}<1.5\times10^{39}$\lunits (Palumbo \etal 1985).
The \rosat PSPC spectrum of the M51 nucleus is fitted with a thermal
plasma of kT$\sim$0.4 keV (Marston \etal 1995;Read \etal 1997)
indicating that the AGN does not dominate the nuclear soft X-ray
emission. 
The \rosat PSPC observations revealed extended emission (Ehle \etal 1995).  
At the hard energies the \ginga\, data are fitted with a
photon index of $\Gamma=1.4$ and an X-ray luminosity of
$L_{2-20 keV}=(1.2\pm0.6)\times10^{41}$ \lunits \,from a $\sim$ 1 deg$^2$ field
containing M51 (Makishima \etal 1990). The data are also fitted with
a kT=7 keV thermal bremsstrahlung model plus a power-law with $\Gamma=1.6$
absorbed by a column of $4\times10^{23}$ \cunits.
Terashima \etal (1998) analysed \asca data and found extended emission
from M51 in the 2-5 keV energy range.
They detected a soft thermal emission represented by either
kT$\sim$0.4 keV with low iron abundance or two kT plasmas ($\sim0.3$
keV  \, \rm and \, $\sim0.8$ keV). 
The hard emission is represented by a power-law with
$\Gamma\sim1.4 \, {\rm{and}} \sim1$ respectively.
The fact that extended emission is observed in the 2-5 keV band, clearly
suggests that the AGN emission contributes only to the hardest end of
the \asca spectrum and is either suppressed at the softer energies or
is not the dominant contributor of the energy output at these energies.
%That indicates that the AGN is heavily obscured or even Compton thick.

Our best fit model for M51 is the composite model
with $kT\sim$0.7 keV and a spectral slope of $\sim1.7$. However we
obtained an upper limit for the column density of 1.6$\times 10^{21}$
\cunits. This is not sufficient to obscure the AGN X-ray emission up to
$\sim$ 5 keV, as it is indicated by the analysis of the brightness
profile of M51. 
The emission could be interpreted as the superposition of thermal emission at soft
energies (below 2 keV), emission from low-mass X-ray binaries (LMXBs),
which contribution dominates at the 2-5 keV band and a power-law
component from the AGN, which is revealed above 5 keV. In this case
the active nuclei should be obscured by a column of
$>5\times10^{23}$\cunits.
However, because the AGN and the LMXBs, both show a power-law spectrum of 
$\Gamma\sim 1.7-1.8$, it is possible both components to be fitted by the
power-law model with no need of excess absorption. 
On the basis of its $[OIII]\lambda5007$ flux, M51 is a Compton thick candidate.
We note here that we also tried a Compton thick model, with either 2
Raymond-Smith models and a reflected continuum (pexrav model), 
or a Raymond-Smith
model, a power-law model to account for the possible LMXBs
contribution and the reflection component. 
However, although we
obtained good fits the reflection component did not contribute in the
\asca band leaving the 
 possibility where the hard X-ray emission is dominated by 
 a scattered component more plausible in this band.

\section{Discussion}

\subsection{Are most of our objects Compton thick ?}
A couple of our objects show no evidence for intrinsic absorption.
%The X-ray data alone cannot discriminate between the Compton thick and
%Compton thin scenarios.
 We note that in the case of poor 
 photon statistics a Compton thick object could 
 be misidentified for a low-$N_H$, steep spectrum 
 type-1 AGN, especially if the 
 steep scattered emission dominates over the reflected component.
 %this may have been the case in NGC3147 (Ptak et al. 1997).  
It is thus possible that \asca only 'sees' the scattered component 
Therefore, the lack of intrinsic absorption in some of our objects  may 
 indicate that these objects are Compton thick.

Further clues on whether these AGN are Compton thick can be given by studying
the isotropic properties of the galaxy. In the case of an AGN as
isotropic emission we consider the
infrared (IR) and  the hard X-ray emission 
(in the case of Compton thin absorption) both been able to penetrate
the torus,
and the $[OIII]\lambda5007$ line emission produced in the
narrow line region, and thus free of viewing angle effects.
Indeed, Alonso-Herrero \etal (1997) showed that the ratio of the 2-10
keV flux to the $[OIII]\lambda5007$ and to the IR flux are comparable
for obscured and unobscured AGNs.
The advantage of studying isotropic properties then, is that they act as an
indicator of the strength of the nuclear source.
 Maiolino et al. (1998) have proposed  that   
 the measurement of the observed hard
X--ray flux (2-10 keV) against the $[OIII]\lambda5007$ flux, 
is indeed a powerful diagnostic. Moreover,  although the 
line is emitted on the Narrow Line Region (NLR) scales, 
 the host galaxy disk
might obscure part of the NLR and should be corrected for the
extinction deduced from the Balmer decrement (Maiolino $\&$ Rieke 1995).
The corrected $[OIII]\lambda5007$ flux is given by the following
relation (Bassani \etal 1999):
\begin{equation}
F_{[OIII]cor}=F_{[OIII]obs} \times
[(H_{\alpha}/H_{\beta})/(H_{\alpha}/H_{\beta})_0]^{2.94}
\end{equation}
Assuming an intrinsic Balmer decrement
$(H_{\alpha}/H_{\beta})_0=3$.
All the well studied Seyfert 1 galaxy have  $f_{HX}/f_{[OIII]}\geq1$ (Maiolino
1998). An absorption of less than a few times $10^{23}$ \cunits will
lower this ratio by a factor of $\sim5$ with respect to Seyfert 1s.
When $N_H>5\times10^{24}$\cunits then the reduction is about two orders of
magnitude.
%Alonso-Herrero \etal 1997 showed that the mean of the ratio
%log$f_{HX}/f_{[OIII]}$ is 1.88 for a sample of Seyfert 1s and PG
%quasars and 1.81 for Seyfert 2s. In the latter case the authors
%corrected the X-ray flux for attenuation. However the authors did not
%correct the [OIII]$\lambda$5007 flux for the extinction in the
%NLR. Such a correction would lower this value.
The flux ratios are presented in table 7. For each object we use the
flux derived using the most plausible model.
 Of course we have to be cautious as there may be some 
 limitations on the use of the  $[OIII]\lambda5007$
flux ratio as an indicator of the unobscured X-ray emission.
 Indeed, in cases where the ionisation cone
axis lies perpendicular or close to the minor axis there might be lack
of ionised gas. Then the
$[OIII]\lambda5007$ flux may not be a good indicator of the nuclear
strength.
However the disk height for a typical spiral galaxy is of the order
of $\sim100$ pc whereas the size of the region where the bulk of the  
$[OIII]\lambda5007$ flux is produced is of a similar size. Thus 
 it is most likely that the relative orientation of the nuclear accretion
to the host galaxy disk will not significantly affect the $[OIII]\lambda5007$ 
flux. Another caveat that we should take into account is whether 
 the torus hides the innermost regions of the NLR from our view and 
 thus the observed [OIII]$\lambda$5007 is lower than 
 the actual flux produced. 
 Again the small inferred size of the torus (a few pc) relative to the size 
 of the NLR (kpc scale) implies that such an effect does not affect significantly the 
 above ratio. In addition Mulchaey \etal (1994) compared the
properties of samples of Seyfert1 and Seyfert 2 galaxies and showed
that the ratio of the [OIII]$\lambda$5007 emission to the unobscured
hard X-ray emission, is the same for both types of galaxies, indicating
that the [OIII]$\lambda$5007 flux cannot be obscured by the torus.
In the case of the IR emission caution must be taken since 
the IR emission may be
contaminated by starburst emission particularly in the case of low
luminosity AGN where the star formation in the host galaxy dominates
the emission. Multiwavelength observations of our objects have
shown that indeed some of them contain starburst activity (see $\S$5). 
So we choose not to apply this criterion to our data.

\begin{table*}
\caption{} 
\begin{tabular}{lccccccc} \hline 

Name    &$f_{HX}^{\, a} $  &$f_{HX}^{\, b} $ & $f_{[OIII]}^{\, c} $
&$f_{HX}/f_{[OIII]}^{\, d}$ & $f_{HX}/f_{[OIII]}^{\, e}$ & $type^{\, f}$ & $type^{\, g}$ \\ \hline
NGC1167 & 0.04&0.04 & 17  & 0.23 &0.23& ?& CT \\
NGC1667 & 0.08&0.08 & 197 & 0.04 &0.04& ?& CTn  \\
NGC2273 &1.11 &3.65 & 277 & 0.40 &1.32&CTn&CT\\
NGC3079 &0.11 &1.10 & 90  & 0.11 &1.20&CTn &CT \\
NGC3147 &1.63 &1.64 & 9.0 & 18.11 &18.22& CTn& CTn   \\
NGC3486 & 0.05&0.05 & 1.7 & 2.94 &2.94& ?& CT/CTn\\
NGC4698 &1.04 &1.10 & 2.0 & 52 &55& CTn& CTn\\     
M51     &0.11 &0.55 & 150 & 0.07 &0.37& CTn & CT     \\ \hline\
\end{tabular}

$^{a}$ 2-10 keV observed flux in units of $10^{-12}$\funits.  \\
$^{b}$ 2-10 keV unobscured flux in units of $10^{-12}$\funits, using the
most plausible model for each object.  \\
$^{c}$ corrected $[OIII]\lambda5007$ flux in units of $10^{-14}$\funits, taken from
Risaliti \etal 1999. \\
$^{d}$ using the observed 2-10 keV flux. \\ 
$^{e}$ using the absorption corrected 2-10 keV flux for the
most plausible model for each object.  \\
$^{f}$ whether the object is Compton thin (CTn) or Compton thick (CT) after taking into consideration the spectral fitting. \\ 
$^{g}$ whether the object is Compton thin (CTn) or Compton thick (CT) after taking into consideration the $f_{HX}/f_{OIII}$ ratio. \\

\end{table*}

In table 7 it is  noted
whether the object is Compton thin or thick on the basis of both  the X-ray
spectral analysis and the $f_{HX}/f_{[OIII]}$ ratio. It is clear that
in some cases there is a discrepancy between the classifications inferred
from the spectrum and the $f_{HX}/f_{[OIII]}$ ratio.
It is
likely that the X-ray spectrum of Seyfert 2 galaxies and especially
those with high column densities, where the medium and hard X-ray
photons are suppressed, and emission from the host galaxy,
circumnuclear starburst and warm absorbers contributes to the spectrum
significantly. The present (\asca) data does not allow us to
distinguish the contribution of these components.

Summarizing, 
NGC1167, NGC2273, NGC3079 and M51 are Compton thick candidates according to
their $f_{HX}/f_{[OIII]}$ ratio. 
The are best for the three of them --NGC2273, NGC3079 and
M51-- are best fitted with  a highly obscured (N$_H$ $\sim$ 10$^{23}$ - 10$^{24}$\cunits) power-law. However due to large uncertainties in
the derived column density we were not able to distinguish between the 
Compton thin and thick interpretations.
%However the spectral analysis for three
%of them, namely NGC2273, NGC3079 and M51 does not favour this
%interpretation, but instead the data are better fitted by an
%obscured power-law (Compton thin model). 
This 
%discrepancy 
may be due to 
the underlying complexity of the X-ray spectra  as 
shown more clearly in the  case of M51. However the current data,
especially in the case of NGC2273,
are only sufficient to support analysis with simplistic models.
NGC3486 is probably a heavily obscured Sy2.  
Both NGC3147 and NGC4698 have
high $f_{HX}/f_{[OIII]}$ ratios,  comparable to the value obtained
by Alonso-Herrero \etal 1997 if we take into account the extinction in
the NLR, clearly showing that the 2-10 keV X-ray
emission is not affected by absorption.

\subsection{Sy2 galaxies with no absorption in X-rays:
 the case of NGC4698}

%In general our Sy2 galaxies show no or low absorption
%($<10^{22}$\cunits). 
%Investigation of their isotropic properties showed that they are not
%Compton thick objects. 

The most striking example from our low absorption Seyfert 2 is NGC4698. 
The X-ray data do not require absorption and the alternative
hypothesis of Compton thickness 
was ruled out on the basis of the $[OIII]\lambda5007$ flux.
Further evidence for the amount of obscuring
material can be obtained by the iron line
emission. For Compton thin Seyfert 2 galaxies the average
equivalent width value derived from \asca data, is 363$\pm$254 (Gilli \etal 1999), whereas for Compton thick
objects the equivalent width can be well above 1 keV (e.g for
NGC6240 the iron $K_\alpha$ line has an  EW$\sim$1.58 keV).
However, we do not obtain a significant detection of line emission
but can only set a 90 per cent 
upper limit to the equivalent width of
such a feature of  $425$ eV. This value is too low for a Compton thick
object and in that sense rules out the Compton thick interpretation
for this object.

NGC4698 is not the first Seyfert 2 galaxy found which presents no 
intrinsic absorption. 
Ptak et al. (1996) first studied the NGC3147 (which is also included
in our sample) with \asca in the context
of an AGN. They interpreted its X-ray emission as originating in
a Seyfert 1 (no Seyfert classification was available at that moment)
or a heavily obscured Seyfert 2.
Moreover, Bassani et al. showed that NGC7590 has negligible
absorption as well in X-rays, although classified as a Seyfert 2 galaxy.
Using the [OIII] $\lambda$5007 criterion the latter authors ruled out the Compton thick
possibility for both objects and suggested that 
these lack a broad line region so that their appearance as Seyfert 2 is
intrinsic and not due to absorption. Then optical
classification as type 2 object will be explained by the presence of
narrow emission lines only due to lack of the broad line region.

Many models for the formation of broad line clouds 
argue towards a link to the disk (see Collin-Souffrin 1987 and Witt
\etal 1997), thus the existence of a class of
objects without a BLR would set constraints to the conditions under
which a BLR is formed as well as the properties of the disk.
In particular recently Nicastro 2000 presented a model in which a
standard accretion disk 
accreting at low rates is not expected to produce broad line emission lines.
According to this model a vertical disk wind, originating at a
critical distance in the accretion 
disk, is the origin of 
the broad line emission region. The disk wind forms for external
accretion rates higher than a minimum value $\dot{m}_{min}$ below
 which a standard disk is stable. For accretion rates
\.{m}$>\dot{m}_{min}$ the disk is unstable 
and a stabilising, 
co-accreting 
``disk/corona+wind'' system forms. The minimum accretion rate is 
$\dot{m}_{min}\simeq0.3{\it{n(am)^{-1.8}}}$, 
where {\it{n}}=0.06 is the efficiency of the accretion, {\it{a}}=0.1
is the viscosity coefficient 
and {\it{m}} is the mass of the black hole. 
This gives a minimum accretion rate of
$\dot{m}>\dot{m}_{min}\sim$(1-4)$\times10^{-3}$ for {\it{m}} 
in the range $10^6-10^9$ M$_\odot$.
Providing that the accretion rate in NGC4698 is low, this model
explains the absence of broad 
emission lines. 
Assuming that the mass of the black hole residing in NGC4698 is the
typical 10$^6$M$_\odot$ for Seyfert galaxies, 
the X-ray luminosity in the 2-10 keV energy band (2.2$\times10^{40}$
\lunits) is $\sim$ 3 orders 
of magnitude lower than 
the observed X-ray luminosities ($\sim10^{43}$ \lunits) in Seyfert 1
galaxies. This discrepancy could 
easily attributed to a lower accretion rate, which immediately
explains the absence of broad emission lines in the spectrum of NGC4698.

The lack of column density could also be explained by the
presence of a dusty warm absorber. The presence of dust accounts for
the optical obscuration of the broad line region, whereas the lack of
X-ray absorption is due to the ionisation state of the absorber. In such
a case cold absorption is no longer required. 
We note here that when we discuss dusty warm absorber models we should
consider that there are two grain destruction mechanisms that must be
taken into account. Firstly the sublimination of the grains when they
become too hot ($\sim$ 2000 K) and secondly, thermal sputtering, which destroys
the dust once the gas electron temperature reaches $10^6$ K(Draine
$\&$ Salpeter 1979; Laor $\&$ Draine 1993). Assuming that the warm
absorber is photoionised, the gas temperature in a typical warm
absorber where oxygen is highly ionised is only
T$\sim5\times10^4$K and thermal sputtering is negligible.
If collisional ionisation plays a significant role then T$\sim10^6$ K
and the dust will be destroyed. However, Reynolds \etal (1997) showed
that photoionisation dominates the ionisation of the plasma unless
r$\geq$100 pc, where r is the distance of the warm absorber from the
central engine. In this case a dusty warm absorber model is viable.
Komossa \etal (1998) predicted the presence of a Carbon edge at 0.28 keV and
showed that the dusty warm absorber smoothes the oxygen edges making it
difficult to be detected. 
%However, dusty warm absorber fits to real
%data do not show any smoothen in the oxygen edges.
Unfortunately \asca is not sensitive enough
at these soft energies, and we cannot test an actual dusty warm
absorber model to our data.
Since at  energies above 0.3 keV both the dusty warm absorber and the 
warm absorber are expected to imprint similar spectral features in the
spectrum and because the X-ray photons are not affected by the dust, 
we fit our data in the energy range 0.5-10 keV with a warm absorber
model in order to examine the
possibility of the presence of ionised material in the very central region
of NGC4698. Although we obtain a good fit to the data ($\chi^2=244.8$ for 265 d.o.f )the ionisation
parameter of the warm material is unphysically high implying that the
data do not require any kind of absorption.
Further clues on the nature of this galaxy can be given by looking
for short time variability in the 2-10 keV band. 
The data do not show evidence for variability, however we note
that this could be due simply to our limited photon statistics.

An insight into the energy production mechanisms at 
different wavebands can come from the spectral energy distribution
(SED) of the galaxies.
The SED for NGC4698 (fig.2, stars) reveals an unexpected emission
distribution compared to the  median radio-quiet (solid line) SED of Elvis
\etal (1994). For comparison we also show the SED of the archetypal
Compton thick Seyfert 2 galaxy NGC1068 (squares). It is clear that the
NGC4698 SED 
deviates from both the typical AGN and the Compton thick AGN SED. The
latter confirms our analysis that NGC4698 is not a Compton thick
Seyfert 2 galaxy. 

To investigate the differences
we computed the 'optical/X-ray' spectral index (Tananbaum \etal
1974) $\alpha_{ox}$
between 2500$\AA$ and 2 keV, which is defined as:
$\alpha_{ox}$=-0.384 log $[\frac{F_v(2 keV)}{F_v(2500\AA)}]$. 
Since there are no measurements in the ultraviolet we made the
conservative assumption that the there is no UV bump and we
extrapolate from the optical into the UV as a straight line in the
$\nu F_\nu$ space.
Then $\alpha_{ox}\sim$ 2.5. Typically it is found that
$\alpha_{ox}\sim$1.4 for AGNs. It is evident that NGC4698 is relatively weak
in the X-rays. However as has already been shown, there is evidence
that the weakness is
intrinsic and not due to absorption, unless the UV and or X-ray emission vary significantly.
We then computed the IR/HX ratio as defined by Mulchaey \etal 1994
and obtained $\sim$ 1.5. This is an upper limit since the 
flux at
25$\mu$m is an upper limit. However it is clear that the ratio is
comparable to that of Seyfert galaxies (Mulchaey \etal 1994).
The discrepancy between the  SED of NGC4698 and that
of other  AGNs could be explained by either  an excess in the optical emission or 
by a deficit in both the IR and X-ray emission. However caution must be taken in
interpreting the SED, since the measurements have been taken at different
epochs and using different apertures.

%The study of the spectral energy distribution (SED) of the galaxies is
%useful in studying the energy output at different wavebands. 
%For Seyfert 2 galaxies the torus suppresses the UV and
%optical emission from our line of sight and reradiates it in the far infrared.
%Thus the far infrared emission dominates the
%SED. 
%On the other hand in the case of broad line AGNs (quasars and Sy1s)
%the far infrared flux is comparable to the optical and X-ray fluxes. 
%The NGC4698 SED (fig.2, stars) reveal an unexpected emission distribution
%by showing a deficit in flux in the far infrared band such that the
%optical emission dominates the total output. This implies that there
%is not dusty material to be thermalised and consequently no production
%of far infrared radiation. This intriguing result gives support to the 
%the scenario where neither an obscuring medium nor a broad line region
%are required. For comparison we also show the SED for the Compton
%thick galaxy NGC1068 (squares) SED and the median radio-quiet (line) SED of Elvis
%\etal (1994). The difference in the shape of the SED is clear
%with the emission showing a peak at the far-infrared wavelengths due
%to the presence of cold dust in the obscuring medium. 

\begin{figure*}
\includegraphics[height=8.0cm]{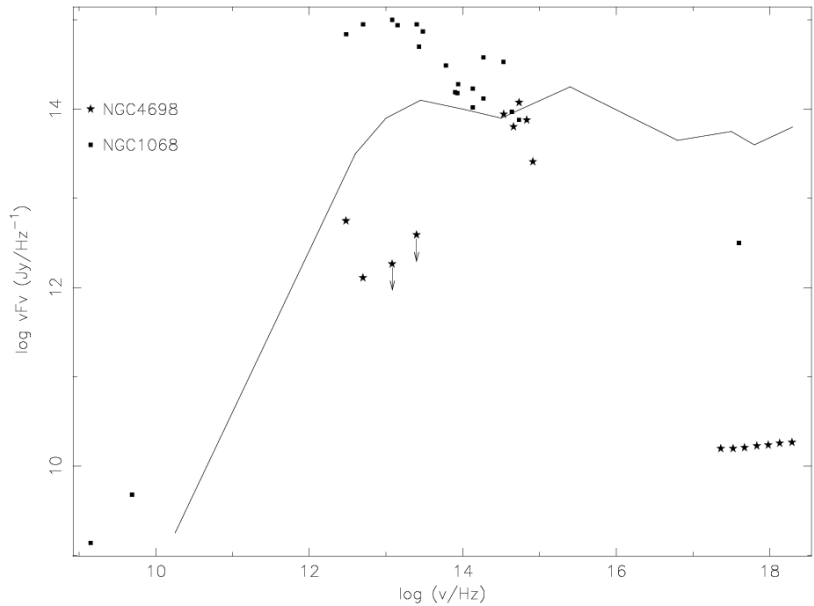}
\caption{The NGC4698 (stars) SED. The fluxes are taken from NED. Where there
are multiple observations, we used data obtained using the smallest
aperture. Specifically, V band magnitude 11.39(64.9'' aperture), I band magnitude
10.54 (36'' aperture) and H band (51.8'' aperture), for comparison we
also show the NGC1068 SED (squares) and the median (line) SED from
Elvis \etal (1994). } 
\end{figure*} 

%\begin{figure*}
%\includegraphics[height=8.0cm]{ngc1068.eps}
%\caption{The average NGC1068 SED. The error bars are standard deviation of the 
%average. }
%\end{figure*} 

\subsection{The distribution of absorbing columns and its 
implications for the X-ray background}

The X-ray background is believed to be produced by the superposition
of discrete sources. In the soft band at a flux limit of
$1\times10^{-15} \rm erg cm^{-2} s^{-1}$ the dominant population are
broad line AGNs (Hasinger \etal 1998).
\asca resolved only  $\sim30\%$ of the 2-10 keV CXB 
into discrete sources (Georgantopoulos \etal 1997) the majority of
which are again broad line AGNs. However broad line AGNs have
power-law spectra with a photon index of $\Gamma\sim1.9$
(Nandra $\&$ Pounds, 1994), which
is significantly softer than that of the CXB in that band ($\sim1.4$,
Gendreau \etal 1995). Thus there must be a large number of undetected
objects which have harder X-ray spectra than the local broad-line
AGNs. This population should be obscured because it does not come up
in the softer energies. 

The deep {\it Chandra} surveys deepened the riddle of the 
 origin of the XRB even further. 
 In the hard 2-10 keV band they probed fluxes at least an order 
 of magnitude deeper than  \asca (Mushotzky et al. 2000)
 albeit with limited number statistics due to the small 
 field-of-view of ACIS onboard {\it Chandra}. 
 A large fraction of the detected sources is associated with 
 QSOs which appear to have steep spectra. 
 Surprisingly, no numerous, clearcut examples of the 
 putative obscured AGN population at high
 redshift have yet been found. 
 Instead, two 'new' populations emerged which are  associated with 
 either early-type galaxies or extremely faint optical counterparts.

All the
current models which try to reconstruct the X-ray background spectra
utilise a population of objects with an underlying spectral index
equivalent of those of unobscured AGNs and a range of obscuring
columns. 
Observational support for such a model comes from the
detection in deep \rosat surveys (Boyle \etal 1995) of a large number of faint
X-ray sources whose optical counterparts are galaxies with narrow lines
only and thus they consider to be that 'obscured ' AGN
population as well as the lack of a population with spectral index
similar to the X-ray background which again indicate that a large
population of heavily obscured AGN should emerge at hard energies.

Our analysis suggests  that simple models cannot describe the
spectrum of the Seyfert 2 galaxies adequately.
Especially in the case of heavily obscured or Compton thick objects,
where the medium and/or hard X-ray emission is suppressed and emission
from the host galaxy, circumnuclear starbursts and/or warm absorbers
contributes significantly and imprints features on the spectrum.
Surprisingly also two of our Seyfert 2s (NGC3147 and NGC4698) do not
show evidence for absorption above the Galactic.
It becomes evident that the complexity of the Seyfert 2 spectrum should
be taken into account when constructing models for the synthesis of the X-ray background.

In addition, latest results from \asca \, and  \bepposax (Pappa
\etal in preparation, Comastri \etal 2000, respectively) show
that there is a population of objects at high redshifts, with broad
lines in optical, which have
high amounts of X-ray absorption.
All the above suggest that the distribution of column densities is
complex, and one cannot use 
a simple recipe for the $N_H$ distribution.
 The compilation of large optically selected or IR selected 
 Seyfert 2 samples, the determination of the X-ray spectrum of Seyfert
2 galaxies as well as the study of obscured
 AGN at high redshift   
will shed new light  on the X-ray background synthesis models.

\section{Conclusions}

We have presented a systematic analysis of 8 bona-fide Seyfert 2
galaxies. We selected sources from the Ho \etal spectroscopic sample
of nearby galaxies. We
included all the brightest $[OIII]\lambda5007$ Seyfert 2 galaxies, for
which X-ray data were or became available.
Our uniform analysis showed that in general our objects show
a complex X-ray spectrum.
On the basis of the $[OIII]\lambda5007$ the Compton thick
possibility was ruled out for two of our low absorption objects,
namely NGC3147 and NGC4698, leaving open questions for the nature of
these objects. We propose that the deficit in absorption maybe either
due to the presence of a dusty warm absorber or due to the lack of
broad line region. In the latter case the Seyfert 2 appearance is intrinsic
and an absorption medium is no longer required. 
%Indeed the study of the SED of NGC4698 
%showed that there is a deficit
%in the far infrared emission, quite untypical even for Seyfert 1 galaxies,
%implying that the torus is not there. The above gives strength to the hypothesis 
%of the lack of a broad line region. 
Furthermore, for NGC2273, spectral analysis favours the scattering
model. However the column density could not be constrained, thus we
could not distinguish between the Compton thin and thick interpretations.  
Yet its $[OIII]\lambda5007$ flux is too low, and favours the Compton thick
interpretation. 
In the case of M51, where the reflection dominated
Compton thick model is not favoured by the data but its 
$f_{HX}/f_{[OIII]}$ is low, we argue that 
the discrepancy may be explained either with a warm scatter model for
the hard X-ray emission or optical or X-ray variability.
Finally we suggest
that the above results may be important in the study of the X-ray
background, since all the current XRB synthesis models utilise a
population of object with an underlying spectral index
equivalent of those of unobscured AGNs and a range of obscuring
columns, whereas our results suggests that there are type 2 objects
with spectrum quite untypical of the one expected and that the X-ray
spectrum is composed by several components, which should be taken into
account when constructing such models..
Discovery of more 
Seyfert 2 galaxies with spectra distinct to the 'nominal' Seyfert 2
spectra will show whether these objects contribute
significantly to the X-ray background or not, while {\it{XMM-Newton}}
observations of Seyfert 2 galaxies will allow us to determine the
spectral components contributing to the total X-ray spectrum and
\chandra observations to deconvolve any extended emission.

\section{Acknowledgments}
The authors would like to thank the referee R. Maiolino for useful 
 comments and suggestions. 
AP wishes to thank Prof. M. Ward for useful discussions and
K. Anagnostou for his useful help throughout the writing of this paper.
This research has made use of data obtained through the High Energy 
Astrophysics Science Archive Research Center Online Service, provided
by the NASA/Goddard Space Flight Center, the LEDAS online service,
provided by the University of Leicester and the NASA/IPAC
Extragalactic Database 
(NED) which is operated by the Jet Propulsion Laboratory, California
Institute of Technology, under contract with the National Aeronautics 
and Space Administration.

\end{document}